\documentclass[preprint,letter,preprintnumbers,amsmath,amssymb,floatfix,nofootinbib]{revtex4}

\usepackage{graphicx}
\usepackage{dcolumn}
\usepackage{bm}
\usepackage{kevin}
\usepackage{amssymb}

\def\W{ {\bf W} }

\begin{document}
\title{Thermal photons from heavy ion collisions:\\
                      A spectral function approach}
\author{Kevin Dusling$^a$ and Ismail Zahed$^b$}
\affiliation{(a)Physics Department, Building 510A\\
Brookhaven National Laboratory\\Upton, NY-11973, USA\\
(b) Department of Physics and Astronomy, SUNY Stony Brook, NY 11794}

\date{\today}

\begin{abstract}
We analyze the photon rates from a hadronic gas in equilibrium using chiral reduction
formulas and a density expansion. The chiral reduction is carried to second order in the pion 
density which in principal includes all kinetic processes of the type $X\rightarrow \pi\gamma$ and $X\rightarrow \pi\pi\gamma$.
The resulting rates are encoded in the form of vacuum correlation functions which are amenable to experiment.  
The hadronic rates computed in this work along with the known perturbative QGP rates are integrated over the space-time evolution of a hydrodynamic model tuned to hadronic observables.  The resulting yields are compared to the recent photon and low mass dilepton measurements at the SPS and RHIC.  Predictions for the LHC are made.
\end{abstract}

\maketitle
\tableofcontents

\section{Introduction}

Electromagnetic emissions in ultra-relativistic collisions are thermally dominated in the low to intermediate mass ($M$) and $q_T$ regions.  Thermalization at RHIC has now been established 
from detailed flow measurements of hadrons.  PHENIX has observed
a large dielectron emission in the mass region below the $\rho$ ($M\approx 770$ MeV).
These emissions are much larger than those reached theoretically. The dielectron
excess reported by PHENIX is one of the most dramatic piece of data stemming from RHIC.

The PHENIX collaboration has also recently reported on photon spectra in the intermediate $q_T$ region.  There is one caveat, in that PHENIX did not actually measure real photons but low mass dielectron spectra which was then extrapolated to the photon point.  In this case their finding was an excess of photons at intermediate $q_T$ ($1 \leq q_T$ GeV $\lsim 2.5$) momentum.  In this case the excess was in line with those reached theoretically. 

Our qualitative summary of the above two PHENIX measurements is as follows.  PHENIX has observed an excess of di-electron pairs at low mass.  This excess is consistent with theoretical expectations at intermediate $q_T$ ($1 \leq q_T$ GeV $\lsim 2.5$) momentum.  The excess seen at low momentum $q_T \lsim 1$ GeV is well above any theoretical calculations done to date.

Thermal emissions at low to intermediate invariant mass ($M$) and $q_T$ are involved due to the many 
reaction processes involving hadrons and the strong character of their interactions.  The 
only organizational principles are broken chiral symmetry and gauge invariance, both of
which are difficult to assert in reaction processes with hadrons in general.   If hadrons
thermalize with the pions and nucleons as the only strongly stable constituents,  then there
is a way to systematically organize the electromagnetic emissivities by expanding them
not in terms of processes but rather in terms of final hadronic states.  The emissivities
are then amenable to spectral functions by chiral reduction. These spectral functions
are either tractable from other experiments or amenable to resonance saturation.

In section II, we derive the photon emission rates from a thermal hadronic environment in terms of one
and two pion final states.  We use the chiral reduction formulae to rewrite the rates in terms of spectral functions. In section 
IIA and IIB we compare our rates to some key processes based
on kinetic theory as well as the leading order QGP rates.  In section
IIC we comment on the experimental extrapolation procedure used recently by RHIC to measure
the photon emissivities by extrapolating the dilepton rates to the photon point.

The remaining part of this works consists of comparison with experimental data.  This is done by integrating the rates over the space-time evolution of a hydrodynamic simulation described in section III.  Sections IIIA, B, C show our findings for WA98 and PHENIX.  Section IIID contains our predictions for the LHC.

\section{Hadronic Photon Rates}
\label{sec:hadpho}

For a hadronic gas in thermal equilibrium the number of photons produced per unit four 
volume and unit three momentum can be related to the electromagnetic current-current 
correlation function \cite{Bellac}
\beqa
q^0\frac{dN}{d^3q}=-\frac{\alpha_{em}}{4\pi^2}\,{\bf W}(q)\,,
\eeqa
with $q^2=0$ and
\beqa
{\bf W}(q)=\int d^4x\spc e^{-iq\cdot x}\Tr\left( e^{-({\bf H}-F)/T} {\bf J}^\mu(x) {\bf J}_\mu(0) \right)\,.
\eeqa
In the above expression ${\bf J}_\mu$ is the hadronic part of the electromagnetic current, 
${\bf H}$ is the hadronic Hamiltonian and $F$ is the free energy.  Below the phase transition the trace is carried over stable states with respect to the strong interaction ({\em e.g.} pions and nucleons).
From the spectral representation and symmetry  we can re-express the correlator in terms of 
the absorptive part of the time-ordered correlation function
\beqa
{\bf W}(q)=\frac{2}{1+e^{q^0/T}}\,\Im {\bf W}^F(q)\,,
\eeqa
where
\beqa
{\bf W}^F(q)=i\int d^4x\spc e^{iq\cdot x}\Tr\left(e^{-({\bf H}-F)/T} \mathcal{T}{\bf J}^\mu(x) {\bf J}_\mu(0)\right)\,.
\eeqa
In this work we will consider a heat bath which is nucleon free.  In this case the trace can be expanded as  
\beqa
{\bf W}^F(q)={\bf W}_0+\int d\pi_1 {\bf W}_\pi + \frac{1}{2!}\int d\pi_1 d\pi_2 {\bf W}_{\pi\pi}+\cdots\,,
\eeqa
where $d\pi_i$ are the pion phase space factors given by
\beqa
d\pi_i = \frac{d^3k_i}{(2\pi)^3}\frac{n(E_i)}{2E_i}\,\,.
\eeqa
In the above density expansion we have defined 
\beqa
{\bf W}_0&=&i\int d^4x e^{iq\cdot x}\langle 0\vert \mathcal{T} {\bf J}^\mu(x) {\bf J}_\mu(0)\vert 0\rangle \nn
{\bf W}_\pi &=&i\int d^4x e^{iq\cdot x}\langle \pi^a(k_1)\vert \mathcal{T} {\bf J}^\mu(x) {\bf J}_\mu(0)\vert \pi^a(k_1)\rangle \nn
{\bf W}_{\pi\pi}&=&i\int d^4x e^{iq\cdot x}\langle \pi^a(k_1)\pi^b(k_2)\vert \mathcal{T} {\bf J}^\mu(x) {\bf J}_\mu(0)\vert \pi^a(k_1)\pi^b(k_2)\rangle 
\label{WWW}
\eeqa
where the latin indices $a,b$ are summed over isospin.  These are the first three terms in an expansion in terms of the pion density\footnote{More specifically the dimensionless expansion parameter is $\kappa\approx n_\pi/(2m_\pi f_\pi^2)$ where $n_\pi$ is the pion density.  This corresponds to $\kappa\approx 0.18, 0.30, 0.84$ at temperatures of $120, 140$ and $190$ MeV.  We therefore expect the expansion to be reasonable unless new thresholds open up.}.

The first contribution in (\ref{WWW}) is dominated by the transverse part of the isovector correlator and is fixed entirely by  
the measured electroproduction data.  It vanishes for real photons since the heat bath is
stable against spontaneous photon emission\footnote{It also vanishes for massive photons having $M\leq 2m_\pi$.}.  Therefore ${\bf W}_0=0$ and does not contribute to the photon emissivities.
The next two terms, ${\bf W}_\pi$ and ${\bf W}_{\pi\pi}$, can be reduced to measurable vacuum correlators by the
chiral reduction formulae~\cite{CRF}. The one-pion reduced contribution ${\bf W}_\pi$ is by now standard. Its dominant
contribution involves VV and AA correlators in the vacuum and reads~\cite{Steele:1996su}
\beqa
{\bf W}^F_\pi(q,k)&=&\frac{12}{f_\pi^2}q^2\text{Im} {\bf \Pi}_V(q^2)\nn
&-&\frac{6}{f_\pi^2}(k+q)^2\text{Im} {\bf \Pi}_A \left( (k+q)^2\right) + (q\to -q)\nn
&+&\frac{8}{f_\pi^2}\left( (k\cdot q)^2-m_\pi^2 q^2\right) \text{Im} {\bf \Pi}_V(q^2)\times\text{Re} \Delta_R(k+q)+(q\to-q)\nn
\label{eq:lin_in_meson1}
\eeqa
where $\text{Re}\Delta_R$ is the real part of the retarded pion propagator, 
and ${\bf \Pi}_V$ and ${\bf \Pi}_A$ are the transverse parts of the VV and AA 
correlators.  Their spectral functions are related to both $e^+e^-$ annihilation 
and $\tau$-decay data as was compiled in~\cite{Huang:1995dd}.  

The two-pion reduced contribution ${\bf W}_{\pi\pi}$ is more involved.
Its full unwinding can be found in~\cite{myThesis}.  We quote only the dominant contributions 
\begin{eqnarray}
\frac{1}{f_\pi^4}W^F_{\pi\pi}(q,k_1,k_2) 
&=& \frac{2}{f_\pi^2}\left[g_{\mu\nu}-(2k_1+q)_\mu k_{1\nu}\text{Re}\Delta_R(k_1+q)\right]\text{Im}\mathcal{T}_{\pi\gamma}^{\mu\nu}\left(q,k_2\right)\nn
&+& (q\to -q) + (k_1\to -k_1) + (q,k_1 \to -q,-k_1)\label{eq:2pig}\nn
&+& \frac{1}{f_\pi^2}k_1^\mu (2k_1+q)^\nu \text{Re}\Delta_R(k_1+q)\epsilon^{a3e}\epsilon^{e3g}\text{Im}\mathcal{B}^{ag}_{\mu\nu}(k_1,k_2)\nn
&-& \frac{1}{f_\pi^2}\left[g^{\mu\nu} - (k_1+q)^\mu(2k_1+q)^\nu\text{Re}\Delta_R(k_1+q)\right]\nn&\times&\epsilon^{a3e}\epsilon^{a3f}\text{Im}\mathcal{B}_{\mu\nu}^{ef}(k_1+q,k_2)\nn
&+& \frac{1}{f_\pi^2}(k_1+q)^\mu(k_1+q)^\nu(2k_1+q)^2\left[\text{Re}\Delta_R(k_1+q)\right]^2\nn&\times&\epsilon^{a3e}\epsilon^{a3f}\text{Im}\mathcal{B}_{\mu\nu}^{ef}(k_1+q,k_2) + (k_1\to-k_1)
\label{eq:2B}
\end{eqnarray}
The first term contains the pion-spin averaged $\pi\gamma$ forward scattering amplitude ($\text{Im}\mathcal{T}_{\pi\gamma}^{\mu\nu})$.  For real photons ($q^2=0$) this is 
entirely constrained by photon fusion data by crossing symmetry \cite{Chernyshev:1995pq} while for virtual photons it can be worked out by further chiral reduction as done in \cite{CRF}.  The remaining terms all contain the contribution labeled $\mathcal{B}$ which reads  
\begin{equation}
\mathcal{B}_{\mu\nu}^{ef}(k_1,k_2)\equiv i\int d^4x e^{ik_1 x}\langle\pi^b_{\text{out}}(k_2) | \mathcal{T}\left( {\bf j}_{A\mu}^e(x) {\bf j}_{A\nu}^f(0) \right) | \pi^b_{\text{in}}(k_2)\rangle\,,
\end{equation}
\begin{figure}
\centering
\includegraphics[scale=0.8]{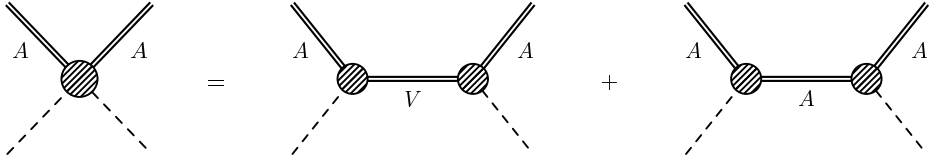}
\caption{Pictorial chiral reduction of $\mathcal{B}$.}
\label{fig:B}
\end{figure}
and is still amenable to further chiral reduction.   The dominant contribution arises 
from the schematic diagrams shown in \Fig{fig:B} resulting in the following spectral
contributions
\begin{eqnarray}
\text{Im}\mathcal{B}_{\mu\nu}^{ef}(k_1,k_2)&=&\frac{2}{f_\pi^2}\delta^{ef}\left[ g_{\mu\nu}(k_1+k_2)^2-(k_1+k_2)_\mu (k_1+k_2)_\nu\right]\text{Im}\Pi_V\left( (k_1+k_2)^2 \right)\nn &+& (k_2\to-k_2)
-\frac{4}{f_\pi^2}\delta^{ef}\left[ g_{\mu\nu}k_1^2-k_{1\mu} k_{1\nu}\right]\text{Im}\Pi_A\left( k_1^2 \right)
\label{eq:Bfinal}
\end{eqnarray}
It is also instructive to look at the above results in the limit where the incoming pions are soft ($k_1,k_2\to 0$).  We then find the simple result that
\begin{eqnarray}
\frac{1}{f_\pi^4}{\bf W}^F_{\pi\pi}(q) = \frac{16}{f_\pi^4}\left( \text{Im}\Pi_A(q^2) - \text{Im}\Pi_V(q^2) \right)
\end{eqnarray}
which reproduces the result of Eletsky and Ioffe\cite{Eletsky:1994rp}.  We should also mention on a more practical level that the dominant contribution to the above expression comes from terms containing $\Pi_V$ and one can therefore neglect terms in \Eq{eq:2B} containing $\mathcal{T}_{\pi\gamma}$ or $\Pi_A$.  The dominant mechanism is therefore the first diagram on the right hand side of \Fig{fig:B}.  This corresponds to the matrix element for $\pi\pi$ scattering via exchange of a vector meson.  The additional terms in \Eq{eq:2B} attach a photon to one of the external lines.

\begin{figure}
\includegraphics[scale=.5]{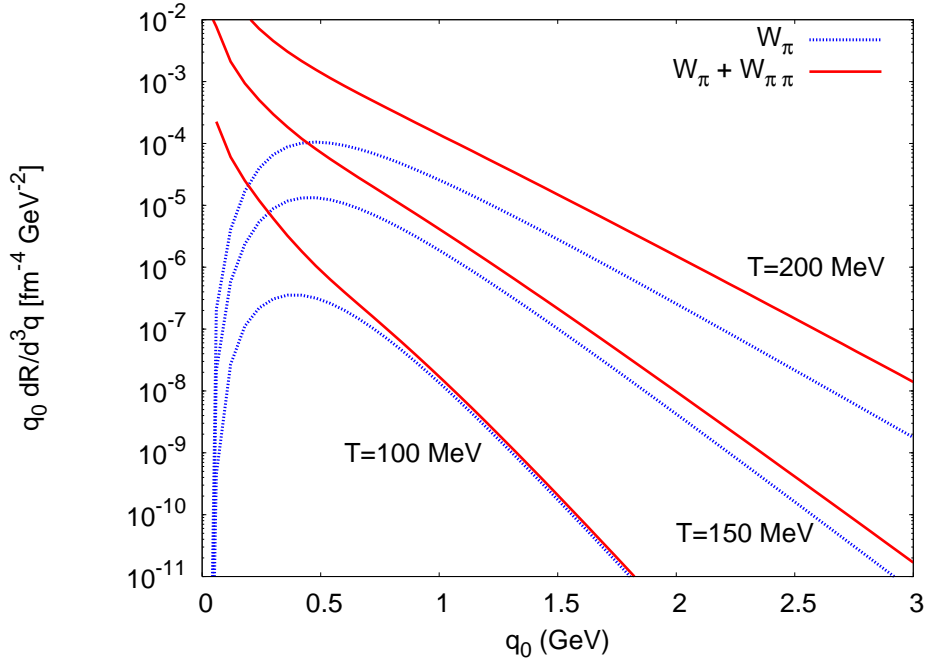}
\caption{Photon emission rates $\W_\pi$ and $\W_{\pi\pi}$ for $T=100, 150, 200$ MeV.}
\label{photoratesCFR}
\end{figure}

In \Fig{photoratesCFR} we show the photon rates stemming from ${\bf W}_\pi$ and ${\bf W}_{\pi\pi}$ for three
temperatures.  Clearly the 2-pion emission rates provide substantial enhancement when the emitted photon is soft due to pion bremsstrahlung.  At the highest temperatures, ${\bf W}_{\pi\pi}$ provides an enhancement even for high energy photons.  But at these temperatures the virial expansion is clearly beyond its limit of applicability.  We stress that the hadronic contributions to ${\bf W}_{\pi\pi}$ involves all hadronic processes in the heat bath with two pions and a photon in the final state.

\subsection{Comparison to Kinetic Theory}

One of the nice features of the above spectral function approach is that one does not need to organize the calculation in terms of the many possible reaction mechanisms, as there is no a priori small expansion parameter in the strongly coupled hadronic phase.
Instead the calculation is organized in a virial-like expansion and all possible reaction channels should be included via the zero temperature spectral densities.  The drawback is that we loose some physical intuition as we tend to discuss photon production from a kinetic theory approach.  In this section we now compare our results with kinetic theory results performed by a number of other groups.

\begin{figure}
\includegraphics[scale=.5]{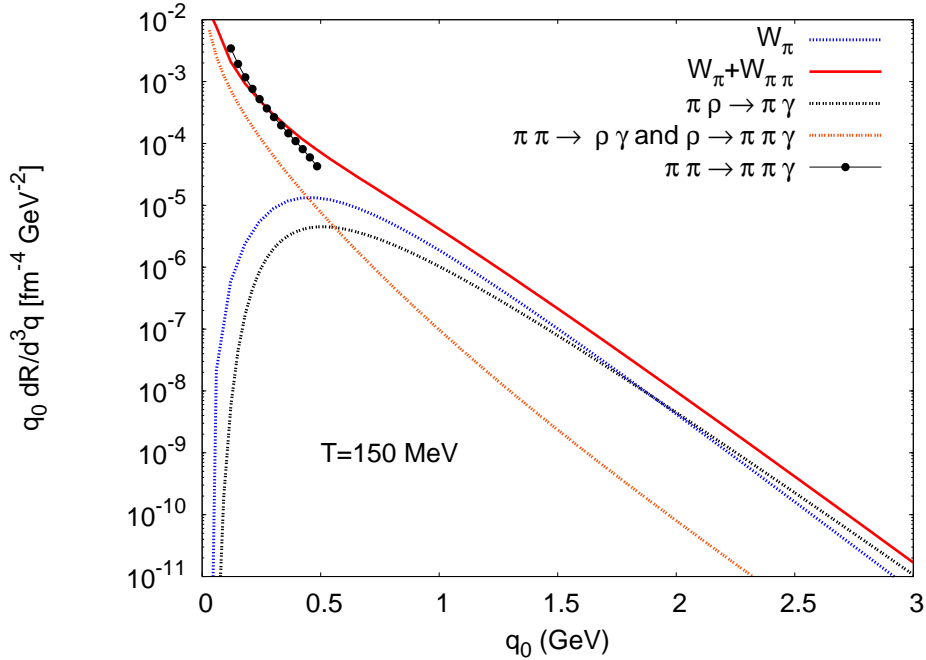}
\caption{Photon rates from spectral functions versus kinetic rates. See text for details.}
\label{fig:compare}
\end{figure}

First let us start with our $\W_\pi$ piece, which was originally worked out in~\cite{Steele:1996su} for both dileptons and photons.  From a kinetic theory standpoint this should include all reactions that contain one pion in the final state.  An example of the dominant contribution in this channel is $\pi\rho\to \pi\gamma$.  In \Fig{fig:compare} we show the one pion piece $\W_\pi$ as well as the rates stemming from the reaction $\pi\rho\to\pi\gamma$ computed in \cite{Turbide:2003si}.  On a qualitative level we find very good agreement considering the rates were derived from two very different approaches.  We should mention that a full SU(3) calculation \cite{Lee:1998um} also includes reactions containing $K$ and $K^*$ ({\em e.g.} $\pi K^*\to K\gamma$). 

We now come to the $\W_{\pi\pi}$ piece which is new to this work.  From a kinetic theory standpoint this should be identified as including all reactions with two pions in the final state and in principal can contain any number of initial state particles.  We expect the reactions $\pi\pi\to\rho\gamma$ and $\rho\to\pi\pi\gamma$ to dominate.  These reactions were calculated in \cite{Turbide:2003si} and are shown in \Fig{fig:compare}.  These two kinetic processes alone cannot explain the observed enhancement that we find in our $\W_{\pi\pi}$ contribution.  Our computed ${\bf W}_{\pi\pi}$ should also contain $2\to 3$ processes such as $\pi\pi\to\pi\pi\gamma$.  This reaction has been worked out beyond the soft photon approximation in \cite{Liu:2007zzw}.  We show a parameterization of their results in the region $0.1 < q_0$ (GeV) $< 0.5$ in \Fig{fig:compare}.  On a qualitative level we see that the pion \brem agrees with our $\W_{\pi\pi}$ contribution.  

We do not claim that the rates encoded in $\W_\pi$ and $\W_{\pi\pi}$ are the end of the story.  In the above comparison we have selected the reaction channels which we expect to be implicitly included in our analysis.  We can conclude that we have a good handle on many of the kinetic theory processes shown above.  The most significant being the $\pi\pi$ \brem encoded in $\W_{\pi\pi}$.  There are other reactions ({\em e.g.} $\omega\to \pi\gamma$), while not included in our $SU(2)$ analysis, have been shown to have strength at low energy \cite{Turbide:2003si,Kapusta:1991qp}.

\subsection{Comparison to QGP Emission}

There has been great progress in the calculation of the QGP photon rates in perturbative QCD.  We will not go through the full history of these results but instead highlight some key points as they may become relevant from a phenomenological perspective.

\begin{figure}[hbtp]
\includegraphics[scale=.5]{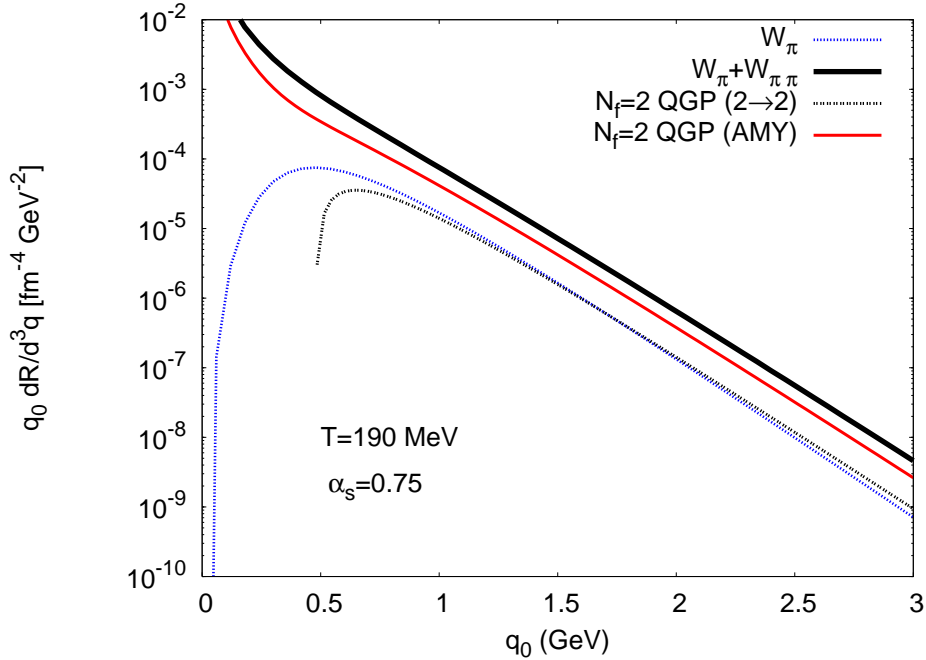}
\caption{Hadronic rates from the spectral function approach versus the perturbative QGP rates.}
\label{fig:hadQGP}
\end{figure}

First off, the born contribution corresponding to $q\overline{q}\to \gamma$ contributes at order $\alpha_s^0$ for dileptons but vanishes at the photon point due to energy momentum conservation.  Therefore, for photons, the first non-trivial contribution will come from one-loop diagrams, $~\mathcal{O}(g^2)$, corresponding to the annihilation $q+\overline{q}\to g\gamma$ and compton $g+q(\overline{q}) \to q(\overline{q}) + \gamma$ processes.  These diagrams contain a logarithmic singularity stemming form the exchange of a soft massless quark.  The singularity is cured by including HTL \cite{Braaten:1989mz} corrections to the quark propagator.  The net result is to give the exchanged quark a thermal mass of order $gT$ and the rates are indeed finite as shown in \cite{Kapusta:1991qp,Baier:1991em}.  We have plotted the photon rate due to $2\to 2$ process in \Fig{fig:hadQGP}.  

It was later realized in \cite{Aurenche:1996is,Aurenche:1996sh,Aurenche:1998nw} that the $2\to 2$ processes of annihilation and Compton scattering are not the only diagrams that contribute at order $g^2$.  It was found that there are a class of diagrams that even though they are naively of higher order are promoted to order $g^2$ due to co-linear singularities.  Two examples of such processes are \brem $(qx\to \gamma q x)$ and annihilation with scattering $(q\overline{q} x\to \gamma g x)$ where $x$ can be either a quark, anti-quark or gluon.  

In order to obtain the leading order result one must resum an infinite set of ladder diagrams which contribute at the same order.  In resumming this infinite set of diagrams one must also take into account the LPM effect which suppresses the rate due to the effects of multiple collisions on the photon production process.  This was demonstrated and evaluated in full in \cite{Arnold:2001ba,Arnold:2001ms,Arnold:2002ja}.  In this work we use the complete leading order rates including the LPM effect as parameterized in \cite{Arnold:2001ms}.  The additional processes included in the full leading order calculation give a large enhancement to the rates as seen in \Fig{fig:hadQGP}.

The leading order QGP rates have one parameter, $\alpha_s$, which should be evaluated at a scale on the order of the temperature.  In this work we evaluate the coupling constant using the two-loop $\beta$ function in $\overline{MS}$ scheme.  In order to asses the uncertainty in the scale choice we have evaluated the yields assuming the coupling runs with two different scales: $\mu=\pi T$ and $\mu=2\pi T$.  At $T=400$ MeV, which is a typical initial temperature in our hydrodynamic evolution we have $\alpha_s(\pi T)=0.38$ versus $\alpha_s(2\pi T)=0.26$.  While at our chosen transition temperature of $T=190$ MeV we have $\alpha_s(\pi T)=0.75$ versus $\alpha_s(2\pi T)=0.40$.  

\begin{figure}[hbtp]
\includegraphics[scale=1]{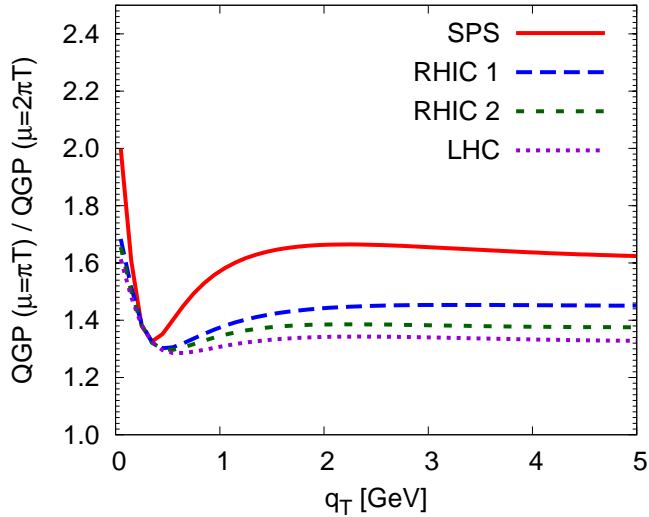}
\caption{Scale dependence of the perturbative QGP yields at SPS, RHIC and the LHC.}
\label{fig:QGPcompare}
\end{figure}

Even though we have yet to discuss the details of the space-time evolution (see \Sect{sec:evolv}) it is still instructive at this point to compare the yields from the QGP phase for various systems in order to see the uncertainty in the scale choice.  In \Fig{fig:QGPcompare} we show the ratio of the QGP yield coming from collisions at SPS, RHIC and the LHC at the two scale choices.  We will focus on the results above $q_\perp \approx 1$ GeV since the perturbative rate calculations breakdown below this scale as we will discuss in a moment.  In general we find an enhancement in the yields by a factor of $\approx 60$\% at the SPS and $\approx 30$\% at the LHC by decreasing the scale from $2\pi T$ to $\pi T$.  At the LHC there is a smaller uncertainty in the scale choice since the temperatures in the system are much higher yielding smaller running coupling effects.  This analysis shows that at RHIC there will be a systematic uncertainty in the yields on the order of $\approx 40$\% due to the scale choice.  With this in mind we have chosen to fix the scale to $\mu=\pi T$ for the remainder of this work in order to get an upper bound on the photon emissivities from the QGP phase. 

Let us now come to \Fig{fig:hadQGP} where we have compared the Hadronic and quark-gluon plasma rates for a broad range of photon energies.  We have generated the plot at the transition temperature $T=190$ MeV, where both rates have the largest theoretical uncertainties. The QGP rates are plotted with a substantial coupling $\alpha_s=0.75$ which is the highest reached in the simulation.  We should stress that in the transition region one should be extremely skeptical of the rates.  For both the hadronic gas and QGP the expansion parameters have become of order one ($\kappa\sim 0.8$ and $g\sim 3$).  One might argue that the QGP rates can be trusted as long as the argument under the log is larger than one.  This is the case as long as one remains above $q_0\sim 1$ GeV.  It is reassuring to see that the photon rates are in qualitative agreement (within a factor $\approx 2$) as one approaches the transition region from above or below. 

\subsection{Photons versus low mass dileptons}
\label{sec:phvd}

In this section we would like to discuss some general considerations between photons and low mass dileptons as well as the rates for the latter.  The discussion of this section will be relevant for understanding the systematic uncertainty in the recent photon measurements at PHENIX as well as understanding part of the source of the low mass di-electron excess observed at PHENIX.  We will withhold the model comparison to data until \Sect{sec:phenixd}.

The rate of dilepton production per unit four volume and four momenta is given by
\begin{equation}
\label{ee}
\frac{dR}{d^4q}=\frac{-\alpha^2}{3\pi^3 q^2}
\,\left(1+\frac{2m^2_l}{q^2}\right)\left(1-\frac{4m^2_l}{q^2}\right)^{1/2}
\frac{1}{1+e^{ q^0/T}}\text{Im}{\bf W}^F(q)
\end{equation}
for massive dielectrons.  As a result we have
\begin{equation}
\label{eeg}
\frac{dR}{d^4q}=\frac{2\alpha}{3\pi\,q^2}
\,\left(1+\frac{2m^2_l}{q^2}\right)\left(1-\frac{4m^2_l}{q^2}\right)^{1/2}
\,\left(q^0\frac{dN^*}{d^3q}\right)
\end{equation}
which ties the dielectron rate to the {\em virtual} photon rate $N^*$ for spacelike momenta.  In the above expression the correlation function ${\bf W}(q)$ is the same as in \Sect{sec:hadpho} but now evaluated at $q^2 < 0$.  Therefore the spectral function approach used for the rates in the hadronic phase naturally contains the rates for photons and dileptons.  The resulting dilepton rates when including the additional two pion piece ${\bf W}_{\pi\pi}$ can be found in \cite{myThesis} and will be briefly discussed later.

The recent photon measurement reported by PHENIX is actually extracted from low mass 
dielectron data below the two pion threshold.  This is done in order to avoid the large background from hadronic decays.  In order to extract the photon rates the PHENIX collaboration makes use of the following relation between photon and dilepton production
\begin{equation}
\label{eegS}
\frac{dR}{d^4q}=\frac{2\alpha}{3\pi\,q^2}
\,\left(1+\frac{2m^2_l}{q^2}\right)\left(1-\frac{4m^2_l}{q^2}\right)^{1/2}\mathcal{S}
\,\left(q^0\frac {dN}{d^3q}\right)
\end{equation}
In the above expression $\mathcal{S}$ is a process dependent factor.  This process dependent factor goes to $1$ for $q^2\to 0$.  For $\pi_0$ and $\eta$ decays (which is the dominant hadronic background) $\mathcal{S}\to 0$ when $M \geq  M_h$(the mass of the hadron).  By cutting out invariant masses less than the $\pi$ and $\eta$ mass the background from hadronic decays will be largely suppressed.

The assumption taken by the PHENIX experiment is that $\mathcal{S}\approx 1$ when $M\ll q_T$.  The high $q_T$, low mass electron pairs are then taken to the photon point via \Eq{eegS}.  Since our spectral function approach has direct access to both photons and low mass dileptons we can easily determine (in a model independent way) the $\mathcal{S}$ factor for hadronic production processes.  

\begin{figure}[hbtp]
  \vspace{9pt}
  \centerline{\hbox{ \hspace{0.0in}
\includegraphics[scale=.3]{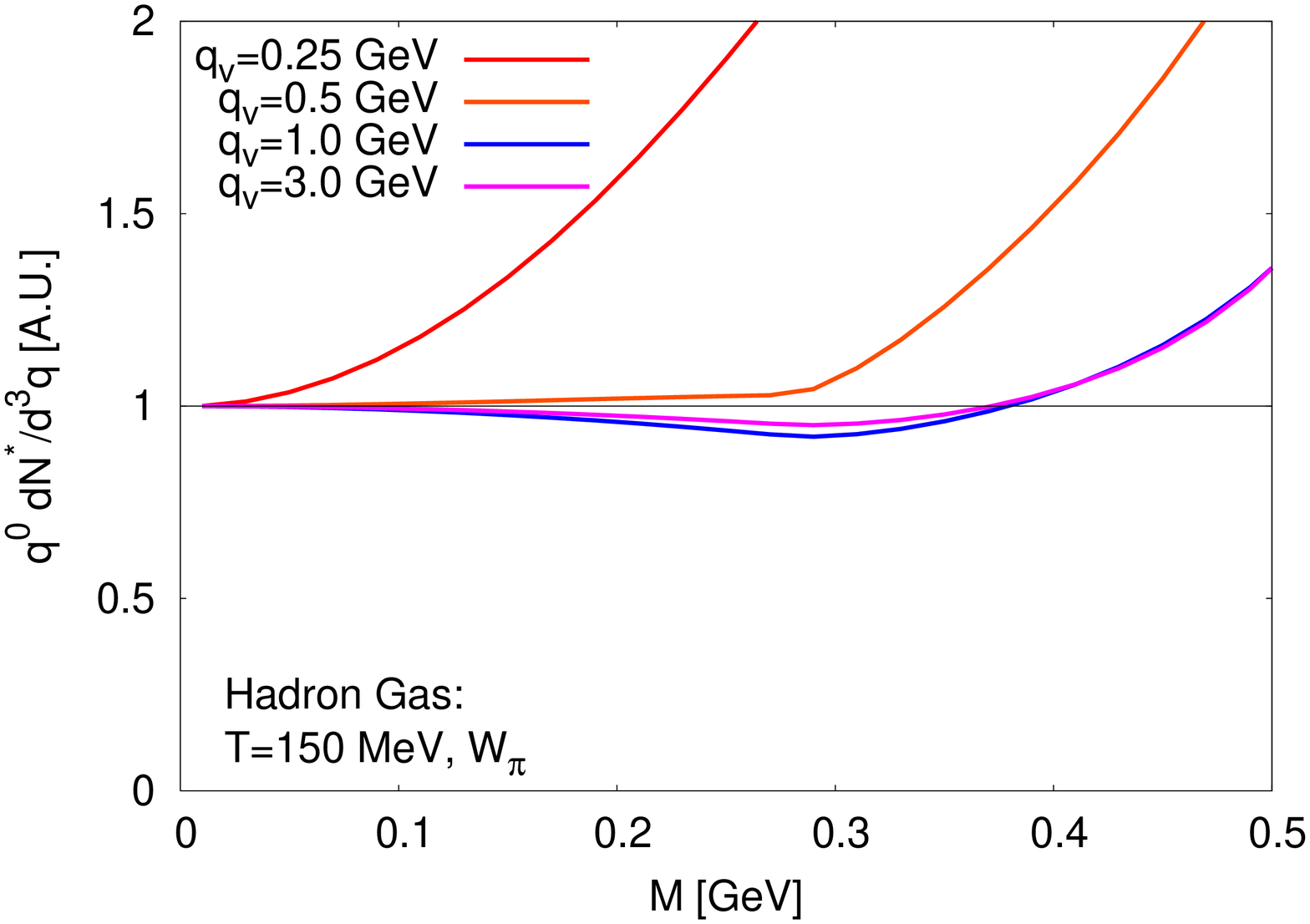}
    \hspace{0.0in}
\includegraphics[scale=.3]{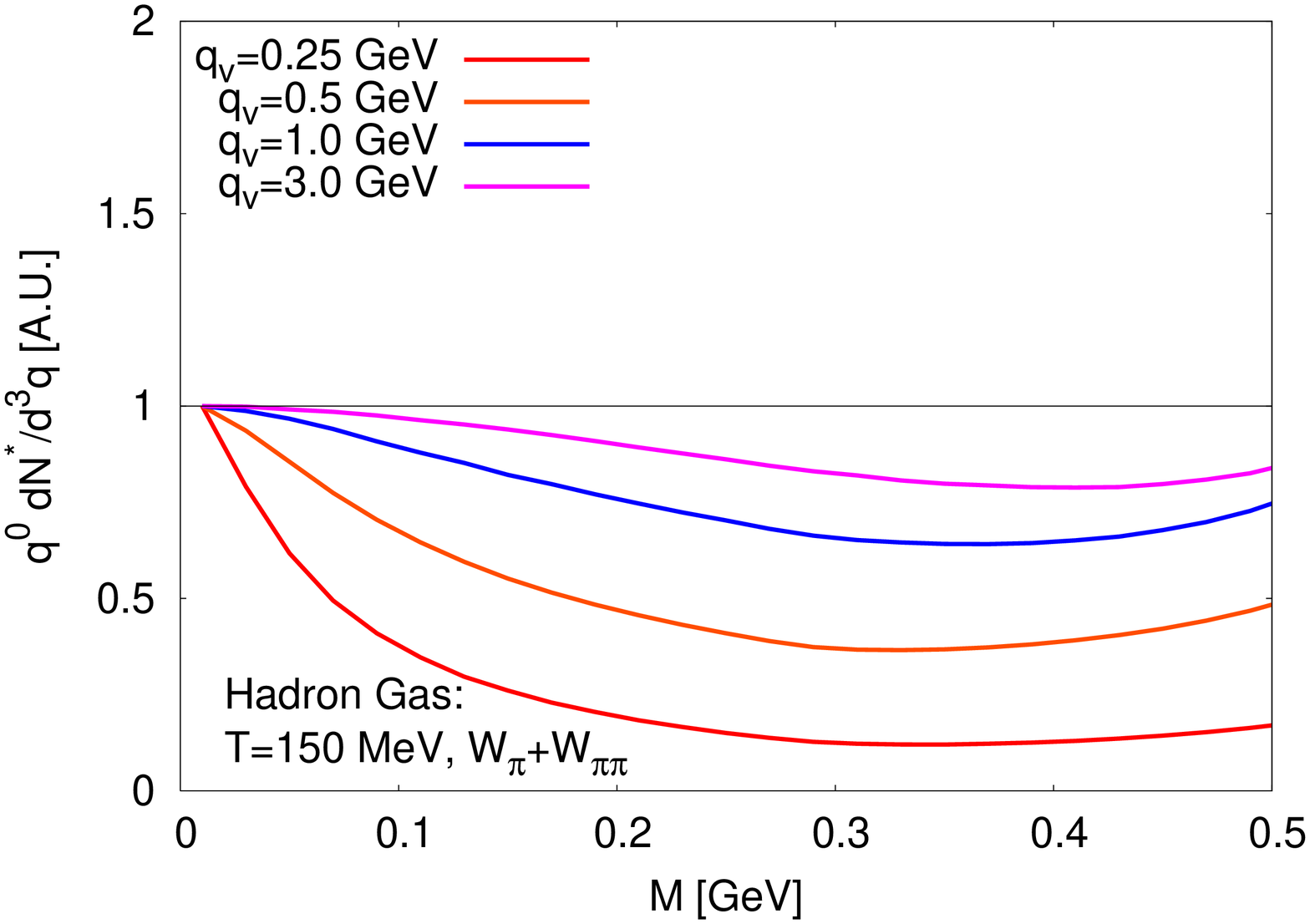}
    }
  }
  \caption{Rate of virtual photons (arbitrary units) from the spectral function approach including: $\W_\pi$ (left) and $\W_\pi + \W_{\pi\pi}$ (right).}
\label{EXTRA}
\end{figure}

In \Fig{EXTRA} we plot (in arbitrary units) the virtual photon rate ($q^0 dN^*/d^3q\equiv \mathcal{S} q^0 dN/d^3q$) where $q^0 dN^*/d^3q$ is the rate of virtual photon production via \Eq{eeg}.  The left figure includes only ${\bf W}_\pi$ while the right figure includes ${\bf W}_\pi+{\bf W}_{\pi\pi}$.  We have fixed the three momentum to $\vec{q}=0.25, 0.5, 1, 3$ GeV.  If our process dependent factor $\mathcal{S}$ was indeed equal to one, we would not see any mass dependence in our plotted $q^0 dN^*/d^3q$.  This is not the case however.  For our leading order ($\W_\pi$) rates at $\vec{q} \geq 0.5$ the factor $\mathcal{S}$ varies by a few percent up to a mass of 300 MeV.  If this was the dominant thermal process then the photon extrapolation made by PHENIX would be satisfactory.  However, when including the more complicated production mechanisms contained in $\W_{\pi\pi}$ the $\mathcal{S}$ factor can change by as much as 30\% at $M=300$ MeV when $\vec{q}=1$ GeV.  Since PHENIX fits the low mass electrons in the $100 \leq M \leq 300$ MeV region we can confidently say that our $\mathcal{S}$ factor will give a discrepancy of $\approx 15$\%.  This leads to a fairly large unconstrained systematic error on the PHENIX extrapolation procedure.  We stress the fact that this error is unconstrained because the $\mathcal{S}$ factor could in principal be very different in the QGP where the production mechanisms may be much different.

Let us conclude this section by a quick discussion on the dilepton rates in the QGP phase.  These will be used in our analysis of the PHENIX data in \Sect{sec:phenix}.  Naively one would expect the leading order Born contribution, $q\overline{q}\to \gamma^*$, to dominate the dilepton mass spectrum.  This breaks down at low mass due to phase space constraints.  For $M\lsim m_{th}$ the next to leading order contribution starts to take over.  All of the same considerations for photons has to be taken into account for dileptons including the HTL resummation, the summation of ladder diagrams and the inclusion of the LPM effect.  This was worked out in full for dileptons in \cite{Aurenche:2002wq}.  For applications we use the parameterization of this result by \cite{Suryanarayana:2007jt,Suryanarayana:2006dd}.  These rates are applicable in the kinematic regions where $q_0 > M$, $q_0 > T$ and $M^2\sim g_s^2 T^2$.  The PHENIX acceptance can fall outside of these regions, so we caution the reader to this additional systematic uncertainty that comes about when comparing to experiment.   To the LPM rates one must also add the $2\to 2$ process worked out in \cite{Thoma:1997dk} which were computed in the approximation $M\lsim T$ and $q_0 \gg T$.

\section{Spectra in Ultra-relativistic Heavy-Ion Collisions}
\label{sec:evolv}

Before a direct comparison with data can be made the photon emission rates of the QGP and Hadronic phases must be integrated over the space-time evolution of the collision.  The collision region is modeled using a relativistic hydrodynamic simulation tuned in order to reproduce hadronic observables.  In this section we discuss the specifics of the model, including the initial conditions and equation of state (EoS),  but leave the technical details to the literature.

The required initial condition for the hydrodynamic evolution is the entropy density, $s(x,y)$, and baryon density, $n_B(x,y)$, in the transverse plane for a given impact parameter\footnote{In the present work we actually do not perform off-central calculations.  Instead the simulations are done in 1+1D where the system size is fixed in order to have the same number or participants.  The resulting yields agree on the 10\% level and are therefore within the overall uncertainty in our hydrodynamic model.} (b) and initial proper time ($\tau_0$).  The entropy and baryon density are proportional to the number of wounded nucleons (participants) which is set by a Glauber model.  More precisely the initial condition is
\beqa
s({\bf r},b)=\frac{C_s}{\tau_0} \frac{dN_{WN}}{d^2{\bf r}}({\bf r};b)\,,\nonumber\\
n_B({\bf r},b)=\frac{C_B}{\tau_0} \frac{dN_{WN}}{d^2{\bf r}}({\bf r};b)\,,
\eeqa
where $C_s\equiv 1/N_{part} dS/dy$ ($C_B\equiv 1/N_{part} dn_B/dy$) is the entropy (net Baryon) density per unit rapidity per participant and is set in order to reproduce the charged particle multiplicity and net proton number respectively.  Once the initial conditions are set the hydrodynamic evolution equations are solved.  The resulting evolution yields the energy density and flow velocity as a function of proper time $\tau$ and transverse coordinate.  The energy density can then be converted to a temperature using the given EoS.  The rates are then integrated over the space-time volume following the same procedure in \cite{Dusling:2006yv,Dusling:2007kh}.

\Tab{tab:param} shows the parameters used for the SPS, RHIC and LHC.  For RHIC we have used two different evolution models (labeled as RHIC 1 and 2) in order to study the effects of the space-time evolution.  We will discuss the specific evolution models as they become relevant in the following sections.  

\begin{table}[h]
\begin{tabular}{l c c c c}
\hline
Parameter & SPS & RHIC 1 & RHIC 2 & LHC \\
\hline\hline
$\sqrt{s_{NN}}$ \mbox{ [A--GeV]} & 17.3 & 200 & 200 & 5500 \\
A & 208 & 197 & 197 & 208 \\
$\sigma_{NN}^{\mbox{in.}} \mbox{ [mb]}$ & 33 & 40 & 40 & 60 \\  
$C_s$ & 8.06 & 20.8 & 20.8 & 42 \\
$C_B$ & 0.191 & 0. & 0. & 0. \\
EoS & BM & Lat & Lat & Lat\\
\hline
\mbox{  Centrality:} & 0-10\% & 0-20\% & 0-20\%  & 0-20\% \\
b [fm] & 3 & 4.5 & 4.5 & 4.8 \\
$N_{\mbox{part}}$ & 340 & 269 & 269 & 293 \\
$\tau_0$ \mbox{ [fm/c]}& 1 & 1 & 0.5 & 0.5 \\
$T({\bf r}_\perp=0,\tau_0) \mbox{ [MeV]}$ & 245 & 336 & 398 & 501 \\
$T_{frzout} \mbox{ [MeV]}$ & 120 & 140 & 160 & 140\\
\hline
\end{tabular}
\caption{Hydrodynamical parameters for: SPS, RHIC and LHC.}
\label{tab:param}
\end{table}

We should stress that we make no attempt to perform global fits to the data.  There are still a number of uncertainties present in the hydrodynamic model which would effect the resulting photon yields. The approach taken in this work is to choose parameters for our evolution model which yields a reasonable description of bulk observables such as the total multiplicity, $p_T$ spectra and elliptic flow.  An example of one uncertainty is the shear viscosity, which when included would require us to re-tune the initial conditions in order to achieve the correct final state multiplicities (as shown in \cite{Luzum:2008cw}).  In addition, viscous corrections will also modify the underlying photon and dilepton rates \cite{Dusling:2008xj,Dusling:2009bc}.  These considerations are all beyond the scope of this work and should be thought of as part of the overall uncertainty in our evolution model.

\subsection{SPS: photons at WA98}

\begin{figure}[hbtp]
  \vspace{9pt}
  \centerline{\hbox{ \hspace{0.0in}
\includegraphics[scale=1]{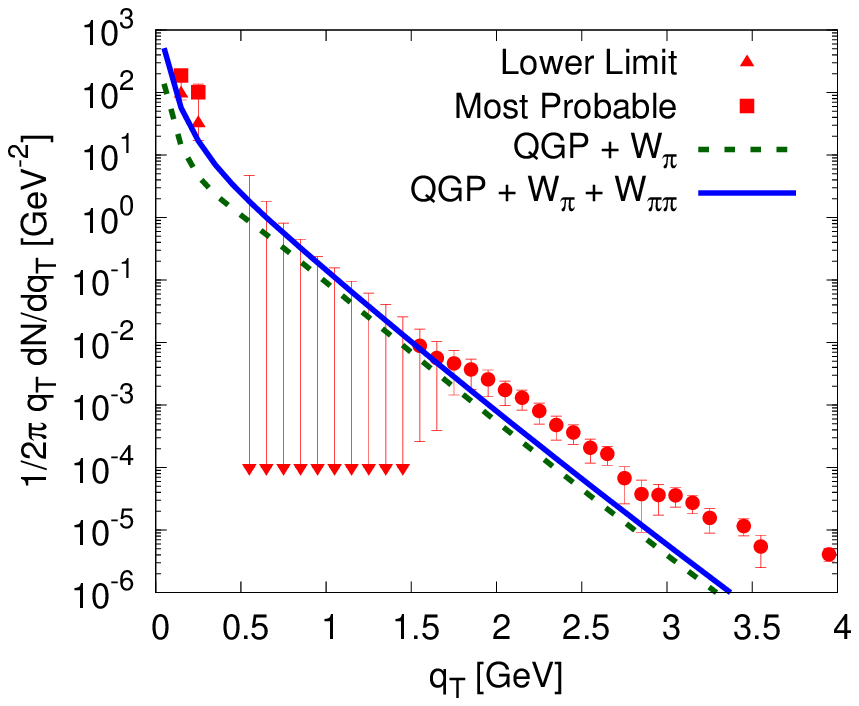}
    \hspace{0.0in}
\includegraphics[scale=1]{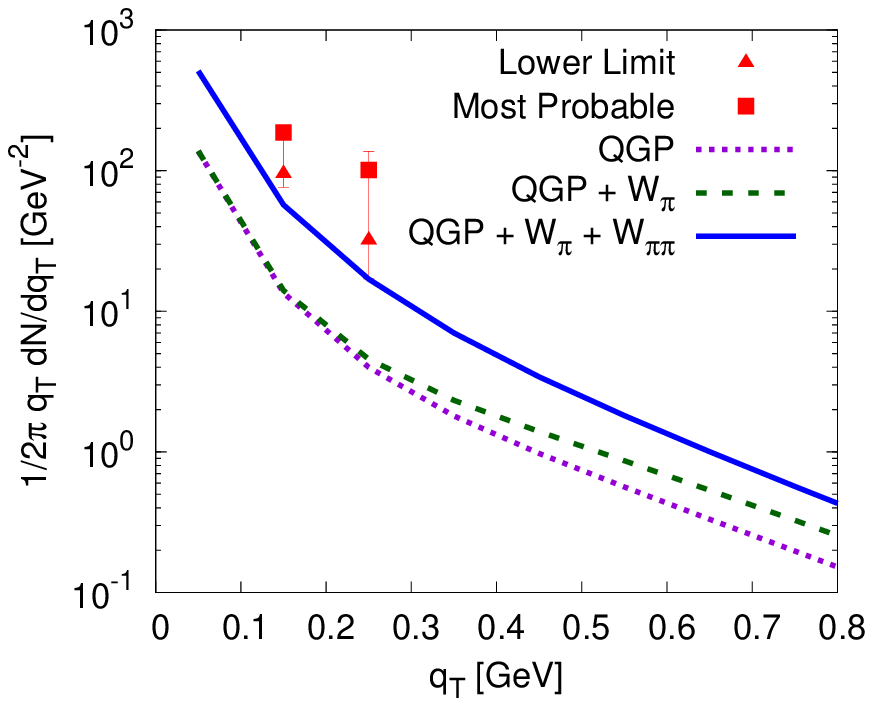}
    }
  }
  \caption{Photon Spectra at the SPS. (Left) Data comparison for $0 \leq q_\perp$ (GeV) $\leq 4$. (Right) Same as left but for $0 \leq q_\perp$ (GeV) $\leq 0.8$.  See text for details.}
  \label{fig:sps}
\end{figure}

The WA98 collaboration has measured direct photons from Pb+Pb collisions at $\sqrt{s}=158$ GeV per nucleon using the statistical subtraction method in \cite{Aggarwal:2000th,Aggarwal:2000ps}.  In this case upper limits could be obtained for photon momentum in the range $0.5 \leq q_\perp \leq 1.5$ and data points obtained above $1.5$ GeV.  In addition the yield of direct photons was also measured in \cite{Aggarwal:2003zy} using direct photon interferometry.  The most probable yield is found by assuming a source size of 6 fm and a lower limit is obtained by assuming a vanishing $R_{out}$.  The data is summarized in the left plot of \Fig{fig:sps}.

\Fig{fig:sps} shows our predicted yields using the SPS hydrodynamic evolution from \Tab{tab:param}.  In this case we have used identical parameters to \cite{Teaney:2001av}.  This work used a bag model EoS having a latent heat of 0.8 GeV/fm$^3$ and was able to reproduce the $p_T$ spectra and elliptic flow at the SPS.  Above 2 GeV one must also include the prompt production which we have left out in this analysis.  This additional contribution was discussed thoroughly in \cite{Turbide:2003si}.  Our results are consistent with the upper limits that were obtained at intermediate momentum.

The right figure of \ref{fig:sps} is the same as the left but rescaled in $q_T$ between 0 and 0.8 GeV.  This figure shows the leading order QGP along with the single pion process $\W_\pi$ are almost an order of magnitude below the data.  By including the additional $\W_{\pi\pi}$ component our rates are now at the lower limit of the data.  In addition one may question how changes to the evolution might effect the results.  It is very probable that one could find a different set of parameters which reproduces the hadronic data at the SPS while also increasing the yields of direct photons.  This type of global analysis is beyond the scope of this work.  It is clear however, that pion \brem as encoded in $\W_{\pi\pi}$ is a necessary component to understanding the direct photons at SPS. 

We should stress that the role played by baryons is large at low momentum \cite{Rapp:1999us}.  While baryons were not included in this work, we can still estimate the enhancement by looking at other calculations.  The work of Steele {\em et al.} \cite{Steele:1997tv,Steele:1999hf} found that baryons, when treated to first order in nucleon density, can increase the rates relative to $\W_\pi$ by an order of magnitude at $q_T\approx 0.5$ GeV.  Based on \Fig{photoratesCFR} we can estimate that the inclusion of nucleons could increase our overall rates (when the $\pi N$ contribution is added to $\W_{\pi\pi}$) by a factor of about two.  This would comfortably keep us within the WA98 data.  

\subsection{RHIC: photons at PHENIX}
\label{sec:phenix}

Let us now come to the recent PHENIX measurements of photons.  For Au+Au collisions at RHIC energies ($\sqrt{s}=200$ GeV) we have chosen to use two different evolution models (called RHIC 1 and 2) in order understand some of our systematic uncertainty stemming from the hydrodyanmic evolution.  In both cases we use a lattice motived EoS \cite{Laine:2006cp} which includes a rapid crossover followed by an interpolation into the hadronic resonance gas phase.  Even though there is no true phase transition, we still must choose a value of $T_c$ where we switch from QGP to hadronic production.  Of course, this choice of $T_c$ will not effect the hydrodynamic evolution which is determined only by the sound speed.  We will fix this to $T_c=190$ MeV which is consistent with lattice calculations at almost physical quark masses \cite{Karsch:2007vw,Cheng:2007jq}.  

The difference between RHIC 1 and 2 is the initial hydrodynamic starting time as well as the freezeout temperature.  For RHIC 1 we have $\tau_0=1$ fm/c and $T_{frzout}=140$ MeV.  In the case of RHIC 2 we have started the hydrodynamic evolution even earlier at $\tau_0=0.5$ fm/c.  In this case, in order to have reasonable agreement with bulk observables one must use a higher freezeout temperature, $T_{frzout}=160$ MeV.  The RHIC 1 parameter set is closer to the one used in \cite{Dusling:2009df,Luzum:2008cw} which found good agreement with the elliptic flow of hadrons.  In the case of RHIC 2 we expect to have a larger contribution coming from the QGP phase and a smaller contribution from the Hadronic phase.  Note that we do not include any out-of-equilibrium contribution (other than the prompt contribution) before our initial starting time $\tau_0$. 

\begin{figure}
\includegraphics[scale=.8]{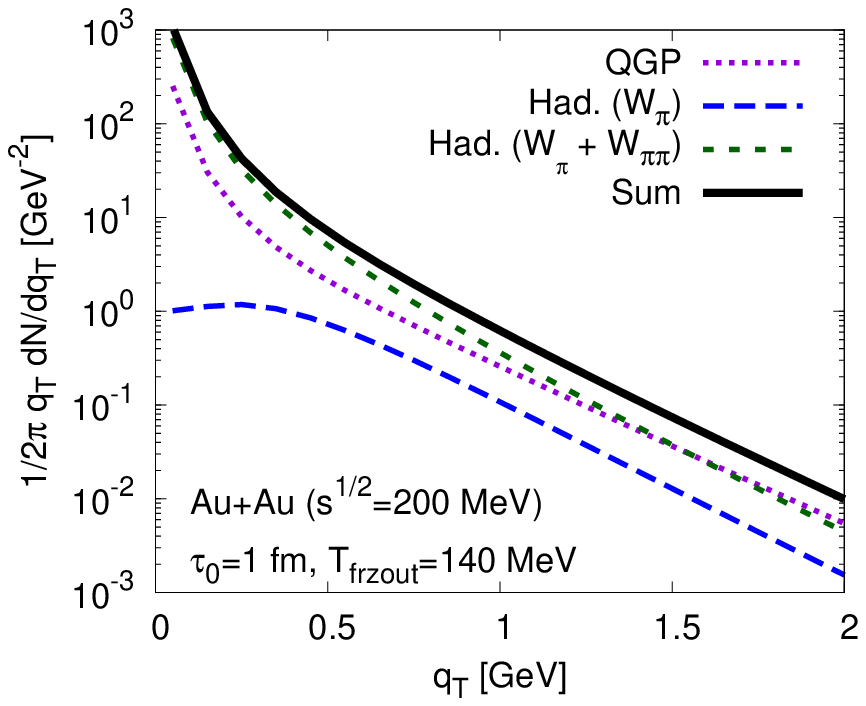}
\includegraphics[scale=.8]{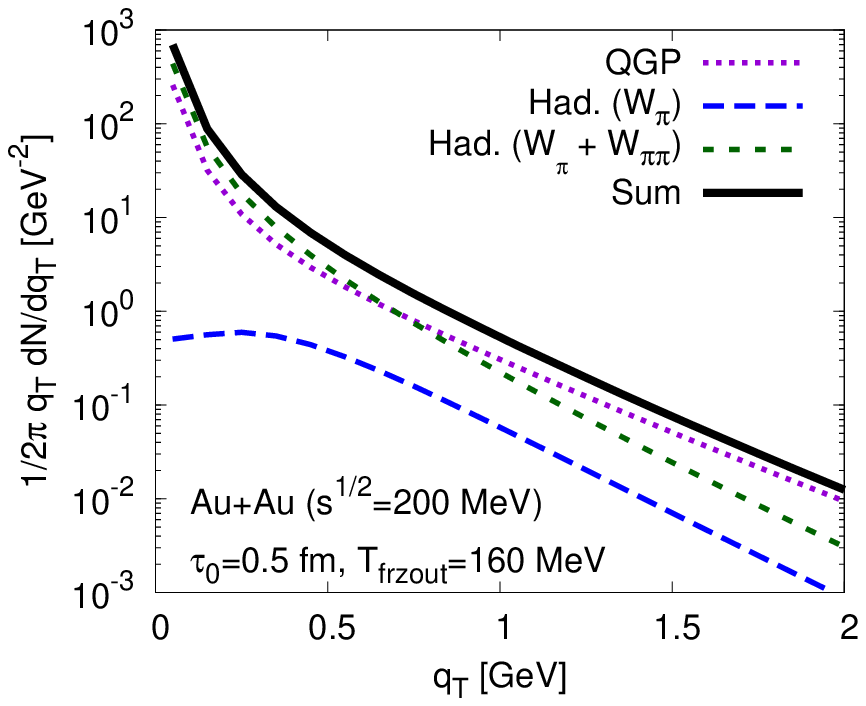}
\caption{Photon spectra from hadronic sources versus the QGP at RHIC.  Left is for the evolution model RHIC 1 and the right is for RHIC 2 (see \Tab{tab:param}).}
\label{figY}
\end{figure}

In \Fig{figY} we show the hydrodynamically evolved photon emissivities using the evolution parameters of \Tab{tab:param} for RHIC 1 (left) and RHIC 2 (right).  The yields from the hadronic gas and QGP cross at $q_\perp \approx 1.5$ and 0.6 for RHIC 1 and RHIC 2 respectively.  The earlier thermalization in the RHIC 2 evolution gives a much larger QGP contribution at high momentum.   

We now come to the recent measurement of direct photons by PHENIX \cite{:2008fqa}.  We will now be slightly hypocritical.  In \Sect{sec:phvd} we have stressed that there will be a large uncertainty in extrapolating the photon signal from low mass dilepton data.  We have demonstrated that for the hadronic gas used in this work that the discrepancy can be as large as $\approx 15$\%.  Regardless we will plot our results for photon production on-top of the PHENIX data keeping in mind that there is at least a 15\% uncertainty in their extraction. 
 
\begin{figure}
\includegraphics[scale=.8]{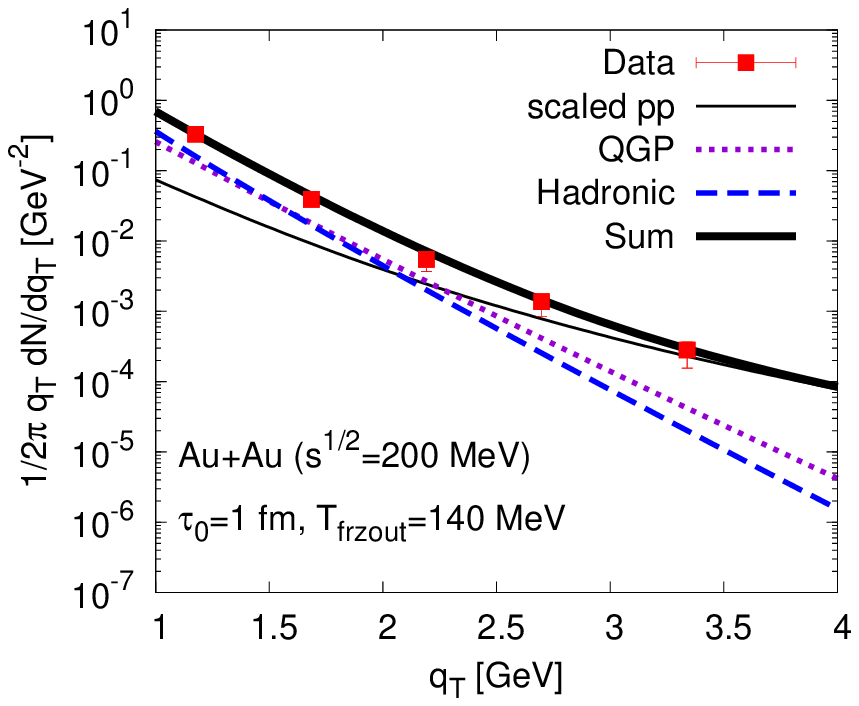}
\includegraphics[scale=.8]{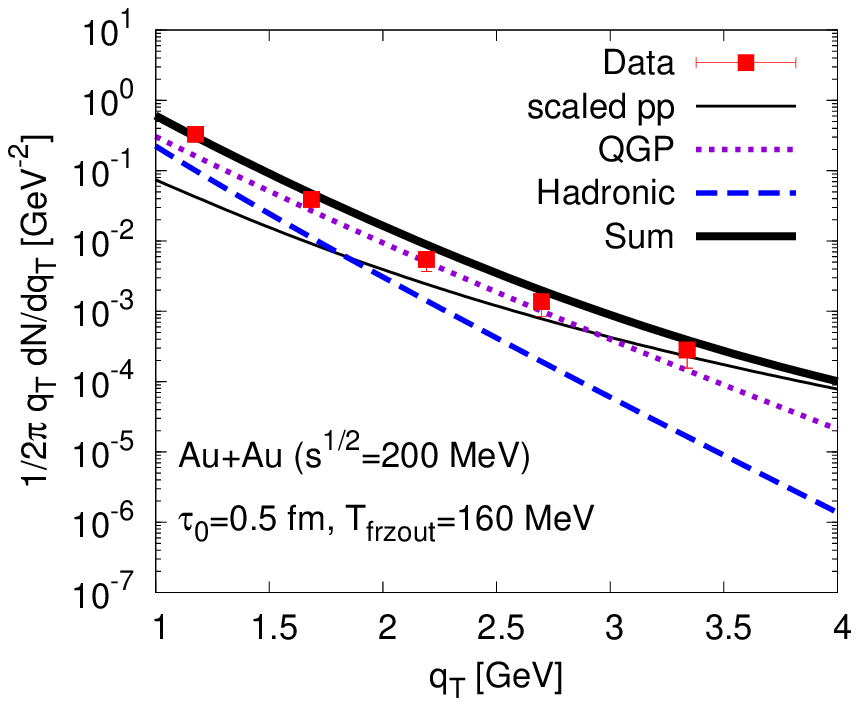}
\caption{Photon spectra at RHIC compared to the recent PHENIX data.  The left plot is the evolution set RHIC 1 and the right is for RHIC 2 (see \Tab{tab:param}).}
\label{RHICDATA}
\end{figure}

Fig.~\ref{RHICDATA} shows the evolved photon rates versus the RHIC data for the two different
hydrodynamical set ups in Table~\ref{tab:param}, (left) is RHIC 1 and (right) is RHIC 2.  We have now also included the prompt photon production which is derived from a fit to the measured photons in pp collisions scaled by the number of binary collisions.  

Our finding is that even though the relative contributions from the different phases are much different the sum remains about the same for the two different evolution models.  In the case of RHIC 1 the relative contributions from the QGP and hadronic are about equal.  For RHIC 2 the QGP has a larger contribution to the overall yield (which grows with increasing $q_\perp$).  The differences in the two models are largest at high $q_\perp$ where the prompt production dominates.  Taking into consideration all of the systematic uncertainties that we have discussed throughout this work it is impossible to discriminate between the two evolution models.  This suggests that there may be a large ambiguity in using photons as an early time probe of the medium.     

\subsection{RHIC: low mass dileptons at PHENIX}
\label{sec:phenixd}

We would now like to revisit our past analysis \cite{Dusling:2007su} of the measurement of dielectron pairs by PHENIX \cite{:2007xw,Toia:2007yr}.  In our prior work we used the leading order Born contribution ($q\overline{q}\to \gamma^*$) as the only reaction present in the QGP phase.  While this is true at high mass the naive perturbative expansion breaks down at low mass which might explain the missing low mass yield from theoretical models \cite{Drees:2009xy}.  We have therefore included in addition to the Born contribution the next to leading order contribution.  The rates used in this analysis were summarized at the end of \Sect{sec:phvd}.  For the analysis of the low mass dileptons we have chosen to use the RHIC 2 evolution model.  Even though this parameter set is for more central collisions $b=4.5$ fm we will make a direct comparison with the min. bias data.  In principal we could perform runs at various centralities and average accordingly in order to make a more direct comparison but this is beyond the scope of this work. 

\begin{figure}
\includegraphics[scale=.8]{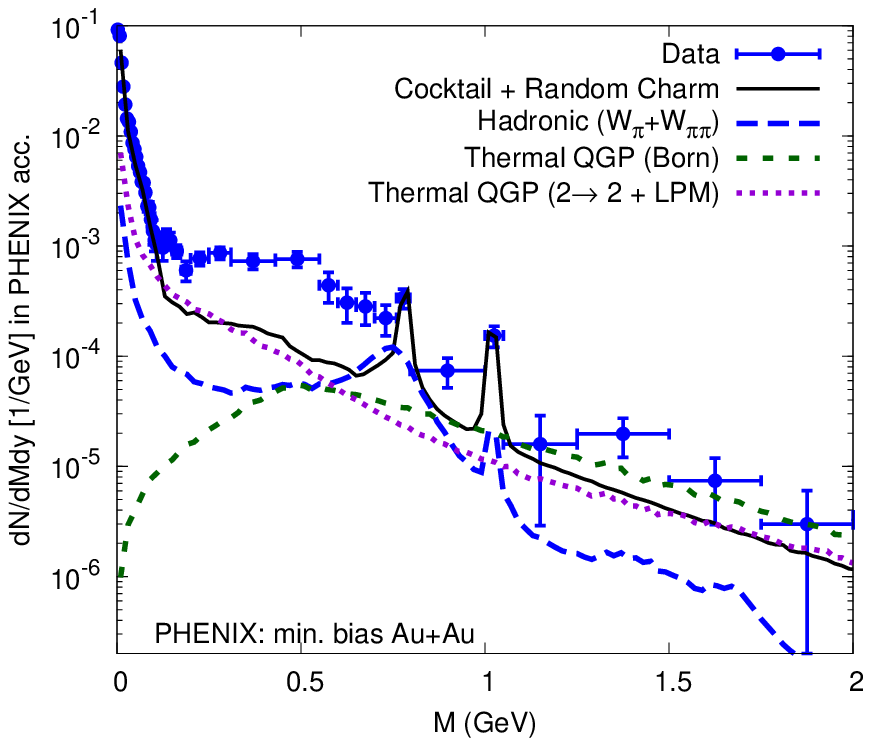}
\includegraphics[scale=.8]{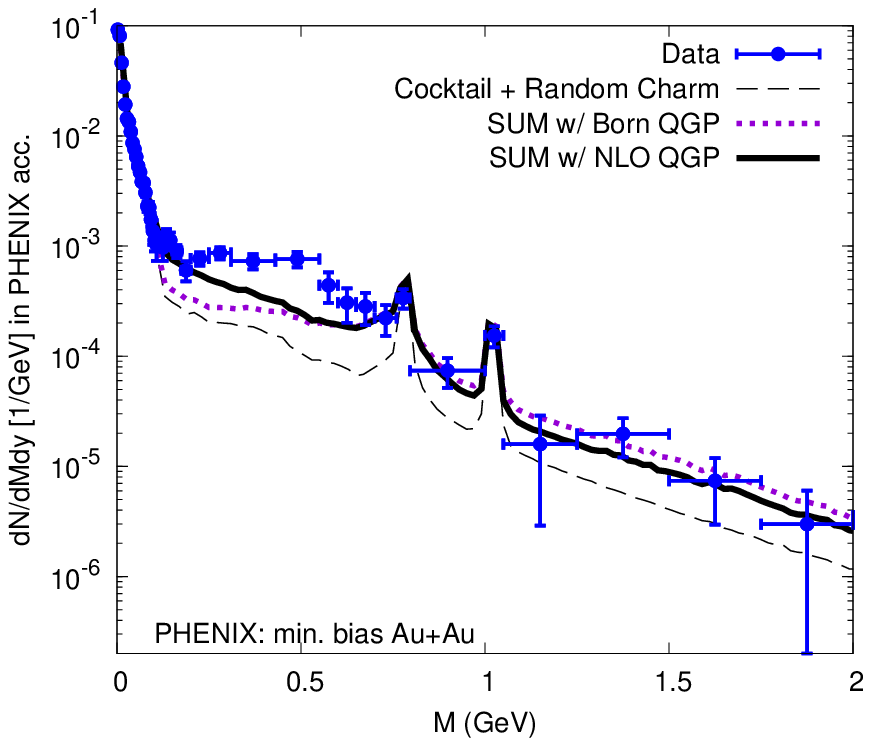}
\caption{Dielectron spectra as measured by PHENIX.  The left shows various individual contributions to the thermal yield.  The right figure shows the total yield using 1) the leading order born contribution and 2) the NLO QGP contribution.}
\label{RHICDATADILEP}
\end{figure}

In the left plot of \Fig{RHICDATADILEP} we show the individual contributions to the low mass dielectron yield.  New to this figure is the inclusion of the $\W_{\pi\pi}$ component in the hadronic phase.  Even though this component added a lot of strength to the low momentum photons the effect is much less dramatic in the case of dileptons.  There is a large enhancement in the dilepton rates at low mass {\em and} low $q_T$ as demonstrated in \cite{myThesis}. By $M\sim 100$ MeV almost all of the enhancement contained in $\W_{\pi\pi}$ is removed by the low $M_T$ cuts of the PHENIX acceptance.  Therefore, to a very good approximation, our $\W_\pi$ makes up the entire hadronic yield at PHENIX after the acceptance cuts.  Also shown in \Fig{RHICDATADILEP} is the qgp Born contribution.  As we can see this contribution diminishes at low mass due to phase space.  However, when we include the next to leading order (NLO) corrections, the QGP yields starts to increase below $2M_{th}$ and actually overtakes our hadronic contribution below $M\approx 500$ MeV.  At high mass (above $\approx 1$ GeV) the NLO rates are suppressed relative to the Born term due to the thermal quark mass\footnote{One could reproduce the NLO rates above $\approx 1$ GeV by simply using the Born rate with massive ($m\sim gT$) quarks.}.

The right plot of \Fig{RHICDATADILEP} shows the total yields which includes summing the cocktail provided by PHENIX along with our thermal hadronic and qgp yields.  We have shown the net yield using only the Born contribution (``Sum w/ Born only'') which is essentially the result from our previous work \cite{Dusling:2007su}.  We also show the yield using the full leading order in $\alpha_s$ result (``Sum w/ NLO QGP'').  The leading order QGP explains the excess in the mass region $100 < M$ (MeV) $<150$.  About 50\% of the thermal qgp production in this mass region is due to the one-loop Compton and annihilation processes.  The other 50\% required to explain the excess comes from the higher-loop processes included in the ladder diagram resummation.  The two-loop piece of this corresponds to the \brem diagrams.  One should not be surprised that the QGP contribution overtakes the hadronic yield at low mass.  It is already known that quark \brem dominates over $\pi$ \brem in the $\vec{q}$ integrated rates \cite{Cleymans:1993jj,Goloviznin:1993rm}. In addition the $q_T$ spectrum of the $\pi$ \brem falls much quicker then that of the quark \brem, which only increases the relative yield of the quark to pion contribution when low $q_T$ cuts are applied.  Larger transverse flow in the hadronic gas alleviates this picture by shuffling strength from low $q_T$ to high $q_T$ but it can not compensate for the differences in the $q_T$ spectrum.   

Even though the addition of the full leading order qgp rates does not explain the excess in the mass region $250 < M$ (MeV) $<550$ it does help in adding some strength to the yield in this mass region.  The mechanism behind the excess in the $250-550$ MeV region is still unclear.  What the authors find striking is the resemblance between the excess seen here and the excess observed in lower energy $C+C$ and $Ca+Ca$ collisions referred to as the ``DLS puzzle''.  The resolution was related to a better theoretical understanding of \brem production in $p+n$ collisions and verified by looking at $p+p$ and $p+d$ control measurements \cite{Zajc:2009je}.  In light of this, it might be useful to have a cocktail available which has been constrained by measured dileptons in $p+d$ or $d+Au$ collisions at RHIC energies.

\subsection{LHC: predictions}

  In this final section we would like to make some predictions for the photon production rate we expect for LHC conditions.  It is very difficult to tune our hydrodynamic model without having any data on the net multiplicities.  Instead we have set our inital condition in order to approximately yield the predicted multiplicity \cite{Kharzeev:2004if,Albacete:2007sm}.  In going from RHIC to LHC the inelastic nucleon-nucleon cross section, which enters into the wounded nucleon profiles has increased from 40 mb to 60 mb.  We have also doubled the initial entropy density.  Our final parameter set is summarized in \Tab{tab:param} and is in qualitative agreement with the hydrodynamic models used in \cite{Niemi:2008ta,Luzum:2009sb}.

\begin{figure}
\includegraphics[scale=.8]{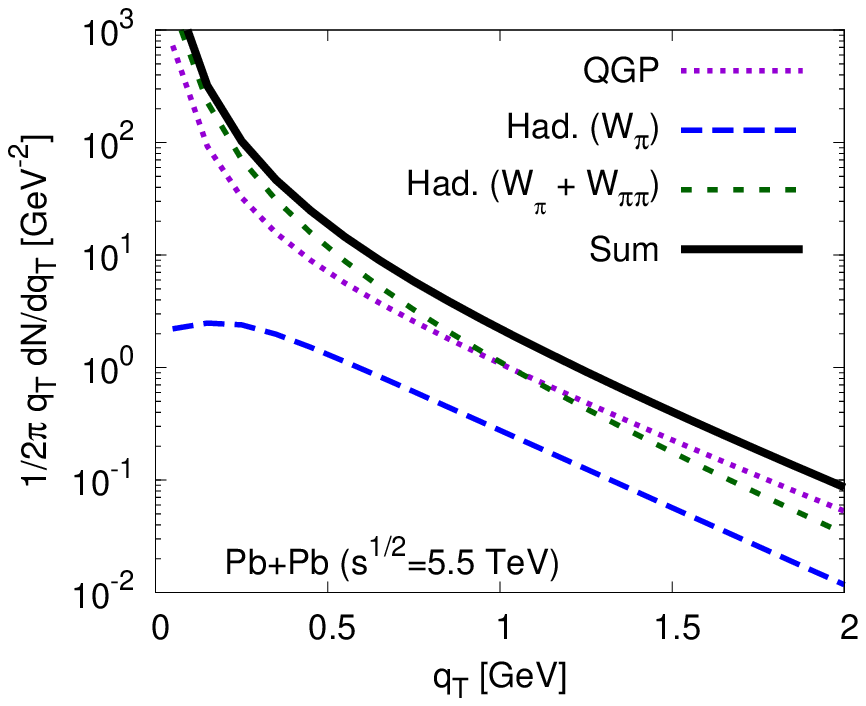}
\includegraphics[scale=.8]{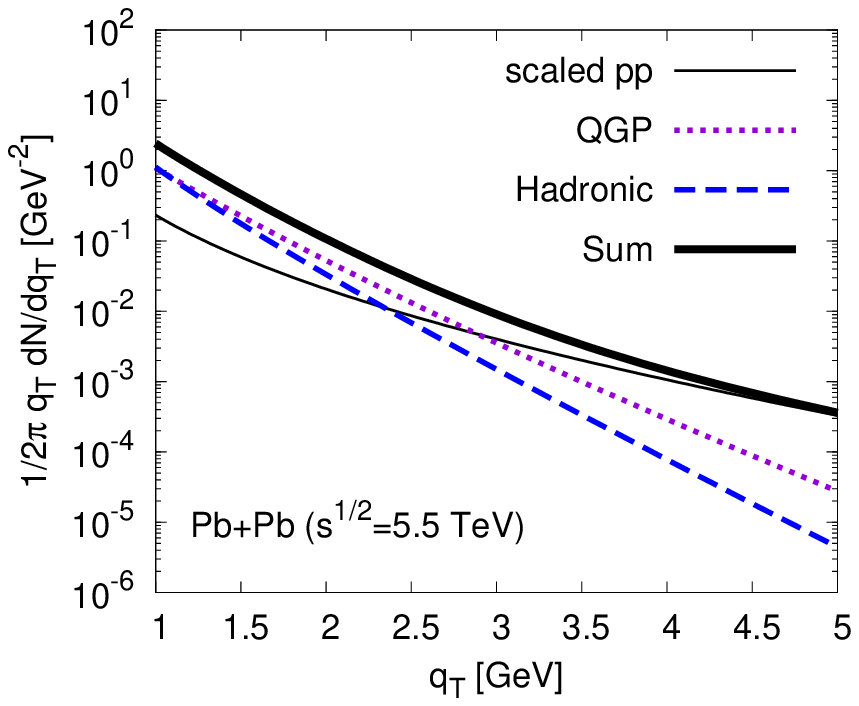}
\caption{Prediction for photon yields from Pb+Pb collisions at $\sqrt{s}=5.5$ TeV.  Left: Relative contribution from $\W_\pi$, $\W_{\pi\pi}$ and the QGP phase for $0 \leq q_T$ (GeV) $\leq 2$.  Right: Relative contributions from Hadronic ($\W_\pi+\W_{\pi\pi}$), QGP and prompt production for $1 \leq q_T$ (GeV) $\leq 5$. }
\label{fig:lhc}
\end{figure}

In Fig.~\ref{fig:lhc} we display our projected photon emissivities at LHC.  The hadronic and qgp yields cross around $q_T\approx 1$ GeV with the hadronic gas dominating at low momentum.  We have also included predictions for the prompt photon production in p+p at LHC energies \cite{Gordon:1993qc} scaled to Pb+Pb collisions. The prompt production starts to dominate the yield above $q_T \approx 2.5-3.0$ GeV similar to what was found at RHIC. 

\section{Conclusions}

In this work we have provided an analysis of the photon emissivities over a large range of collision energies ranging from the SPS ($\sqrt{s}=158$ GeV) to RHIC ($\sqrt{s}=200$ GeV) and to predictions at the LHC ($\sqrt{s}=1.5$ TeV).  The analysis incorporates both hadronic and QGP rates integrated over an underlying hydrodynamic evolution tuned to bulk observables.  In this section we would like to summarize the main conclusions of our work.

\begin{itemize}
\item We have extended the chiral reduction approach to include terms to second order in pion density.  The new addition of this work is inclusion of processes of the type $X\to \pi \pi \gamma$.  The use of a spectral function approach allows us to treat both dilepton and photon production on an equal footing and incorporate broken chiral symmetry in a systematic fashion.  We stress that our hadronic rates are parameter free and are completely constrained by $\tau$ decay, electroproduction and Compton scattering data.

\item Using our hadronic rates, we have tested the low mass extrapolation used by PHENIX and found that we may expect a systematic error in their procedure of at least 15\%.

\item We have found that the pion \brem process, encoded in $\W_{\pi\pi}$, leads to a considerable photon enhancement in the low energy region.  These processes are a necessary ingredient in a quantitative understanding of the low energy ($q_\perp \lsim 500$ MeV) photon data as demonstrated by the recent WA98 measurements (see \Fig{fig:sps}). 

\item The hadronic and LO qgp photon rates are able to adequately describe the RHIC data.  We have found that two different evolution models, which have different relative yields from the QGP and hadronic phases, are both able to describe the data.  

\item We have re-addressed the low mass dilepton data from PHENIX.  Our findings are 1) the $W_{\pi\pi}$ processes are completely removed by the acceptance cuts. 2)  Inclusion of the NLO dilepton rates from the QGP are able to account for the enhancement found in the $M=100-150$ MeV region. 

\end{itemize}

\section{Acknowledgements}

KD would like to thank Stefan Bathe for useful discussion and Dmitri Peressounko for providing the SPS photon result from the HBT analysis.  KD would also like to thank Werner Vogelsang for providing his prompt photon production calculations.  We are also indebted to Axel Drees for stressing to us the role of the NLO corrections in the analysis of the PHENIX dielectron data.  Finally, we are grateful to Ralph Rapp for his careful reading of our manuscript and making many useful suggestions.  KD is supported by the US-DOE grant DE-AC02-98CH10886.  The work of IZ was supported in part by US DOE grants DE-FG02-88ER40388
and DE-FG03-97ER4014.

\end{document}